\documentclass[preprint]{chjaa}
\usepackage{graphicx,times}
\usepackage{indentfirst}
\usepackage{amsmath}
\usepackage{hyperref}
\begin{document}
\title{Improbability of DUrca process constraints EOS}
%
\volnopage{Vol.0 (200x) No.0, 000--000}
%
\setcounter{page}{1}
\author{Hao Tong \inst{1} \mailto{} \and Qiu He Peng \inst{1}}
\institute{Department of astronomy, Nanjing University, Nanjing, 210093, China \\%
           \email{htong\_\,2005@163.com}%
          }
\date{Received~~2001 month day; accepted~~2001~~month day}
\abstract{According to recent observational and theoretical
progresses, the DUrca process (direct Urca process) may be excluded
from the category of neutron star cooling mechanisms. This result
combined with the latest nuclear symmetry energy experiments, will
provide us an independent way of testing the EOS (equation of state)
for supernuclear density. For example, soft EOSs such as FPS will
probably be excluded. %
\keywords{equation of state---neutrinos---stars: neutron}}
\authorrunning{H. Tong \& Q. H. Peng}
\titlerunning{Improbability of DUrca constraints EOS}
\maketitle
\section{Introduction}
Ever since 1965, MUrca process (Modified Urca process) has become
the standard cooling mechanism of neutron stars (Bahcall, Wolf 1965;
Yakovlev et al. 2001; Page, Geppert, Weber 2006). This has not
changed until the 1980s. Boguta first noted this thing (Boguta
1981), and argued that DUrca process (direct Urca process) is
possible in a relativistic nuclear theory. Lattimer et al. made a
thorough investigation about the nucleon and hyperon DUrca process
(Lattimer et al. 1991; Prakash et al., 1992; Pethick 1992).
According to their researches, the critical density of the DUrca
process is determined by the nuclear symmetry energy. If the central
density of a neutron star is above that density, then it will cool
via the DUrca process, which is much faster than the MUrca process.
While for a given neutron star, its central density is determined by
the EOS (equation of state). This means that any information about
DUrca process will provide us an independent way of knowing
something about the neutron star core EOS. Of course, it is only of
prospective use in the 1990s. But there are dramatic changes
recently.

Tsuruta's group has made systematical comparisons between
observations and theories of neutron star cooling. In their point of
view, nucleon DUrca process as well as kaon ones may already be
excluded (Tsuruta et al. 2002; Tsuruta 2004, Tsuruta 2006). Thus if
one EOS permits DUrca process, then it will probably be excluded.
The improbability of DUrca process provides us an independent way of
testing the EOS. There are also tremendous progresses about  nuclear
symmetry energy (Li, Chen 2005; Chen, Ko, Li 2005; Li, Chen, Ko
2006). This two joined together give us deep insight into the
neutron star core EOS. For example, soft EOS such as FPS may be
excluded. Before looking into the EOS, we will review some details
about the DUrca process.
\section{The DUrca process}
Several minutes after a neutron star's borning, it enters the
neutrino cooling epoch. DUrca process is the simplest neutrino
emission process (Gamow, Schoenberg 1941, Pethick 1992). It is
simply decay of neutrons and successive electron captures.
\begin{equation}
\begin{split}
n \rightarrow p + e^- + \bar{\nu}\\
p + e^- \rightarrow n + \nu.
\end{split}
\end{equation}
Whereas, if another nucleon is present as a bystander particle, this
becomes the traditional MUrca process (Chiu, Salpeter 1964; Bahcall
, Wolf 1965, Yakovlev et al. 2001). Since MUrca involves two
additional fermions, it is 5--6 orders slower than that of DUrca.
MUrca process can proceed without difficulty, conserving both energy
and momentum, of course with the sacrifice of a slower rate. For
DUrca process to occur, there is a minimal proton concentration $x$
in order to meet the momentum conservation condition. We will follow
Lattimer's treatment here.

The momentum of the emitted neutrinos and antineutrinos is of order
$k\,T/c$, where $T$ is the neutron star's internal temperature taken
to be $10^{10}\,K$. While the typical Fermi momentum is of order
$\approx 100\,MeV$. Thus, the momentum conservation condition is
$p_{F_p} + p_{F_e} > p_{F_n}$. Noting that $n_i \propto p_{F_i}^3$,
for a $n\,p\,e$ matter, $n_p = n_e$ as a consequence of charge
neutrality, or $p_{F_p}=p_{F_e}$. So, the momentum conservation
becomes $2\,p_{F_p} > p_{F_n}$, or $n_p > 1/8 n_n$. We define the
proton concentration as $x = \frac{n_p}{n_p + n_n}$. Therefore we
obtain the threshold for DUrca process to proceed $x \ge 1/9$. If
the proton concentration exceed that threshold, a neutron star will
cool via the rapid DUrca process.

For a neutron star, the actual proton concentration is determined by
the microscopic interaction, such as the isospin dependent part of
the three body interaction. Nuclear symmetry energy is well among
the list. Employing a schematic model. The energy per baryon can be
expanded quadratically around the symmetry value $x = 1/2$,
\begin{equation}
\epsilon(n, x) = \epsilon(n,\frac12) + S_v(n)(1 - 2\,x)^2 + \cdots.
\end{equation}
Where $n$ is the number density of baryons, $S_v(n)$ is the bulk
symmetry energy. The above expansion is a good approximation for all
x, at any density (Lattimer et al. 1991, and reference therein).

We are considering a system in $\beta$ equilibrium, the chemical
potentials of the fermions have the relation (Shapiro, Teukolsky
1983)
\begin{equation*}
\mu_e = \mu_n - \mu_p = -\frac{\partial \epsilon}{\partial x}.
\end{equation*}
Where $\mu_i$ stands for the chemical potential of the $i\,th$ Fermi
system. Substitute the above expansion of the energy, we get the
equation which determines the equilibrium proton concentration,
\begin{equation}
\hbar c (3\pi^2 n x)^{1/3} = 4 S_v(n) (1 - 2x).
\end{equation}
We may adopt a power law nuclear symmetry energy,
\begin{equation}
S_v = S_0 \begin{pmatrix}
            \frac{n}{n_s}
          \end{pmatrix}^q.
\end{equation}
Where $S_0$ is bulk symmetry energy at nuclear saturation density
$n_s = 0.16\,fm^{-3}$.

A power law symmetry energy has recently been approved by nuclear
diffusion experiment at subnuclear density (Li, Chen 2005; Chen, Ko,
Li 2005). So we are on the edge to see if there is DUrca process in
the interior of neutron stars. Corresponding to the minimum proton
concentration, the critical density is
\begin{equation}
\frac{n_c}{n_s} =  \begin{bmatrix}
                     1.71 (30\,MeV)/S_0
                   \end{bmatrix}^{1/(q - 1/3)}.
\end{equation}
Where $n_c$ is the critical number density corresponding to setting
$x = x_c = \frac19$. For a conservative consideration
(Li, Chen, Ko 2006),%
\begin{equation}
32\,MeV (n/n_s)^{0.7} < S_v < 32\,MeV (n/n_s)^{1.1}.
\end{equation}
Selected values of $n_c$ are given in table \ref{tq}.
\begin{table}
  \centering
  \caption{DUrca Process Critical Density for Different Values of q.}\label{tq}
\begin{tabular}{llllll}
  \hline
  q & 0.7 & 0.8 & 0.9 & 1.0 & 1.1 \\ \hline
  $n_c/n_s$ & 3.62 & 2.75 & 2.30 & 2.03 & 1.85 \\
  $n_c\,(fm^{-3})$ & 0.58 & 0.44 & 0.37 & 0.32 & 0.30 \\
  \hline
\end{tabular}
\end{table}
Combined with the EOS, we can say whether there is DUrca process in
the interior of  neutron stars. But whether it exists, can only be
inferred from neutron star cooling observations. That is the subject
of the next section.
\section{Improbability of DUrca}
\begin{figure}[t]
\begin{minipage}[t]{0.5\textwidth}
\includegraphics[height=.45\textheight, angle=270]{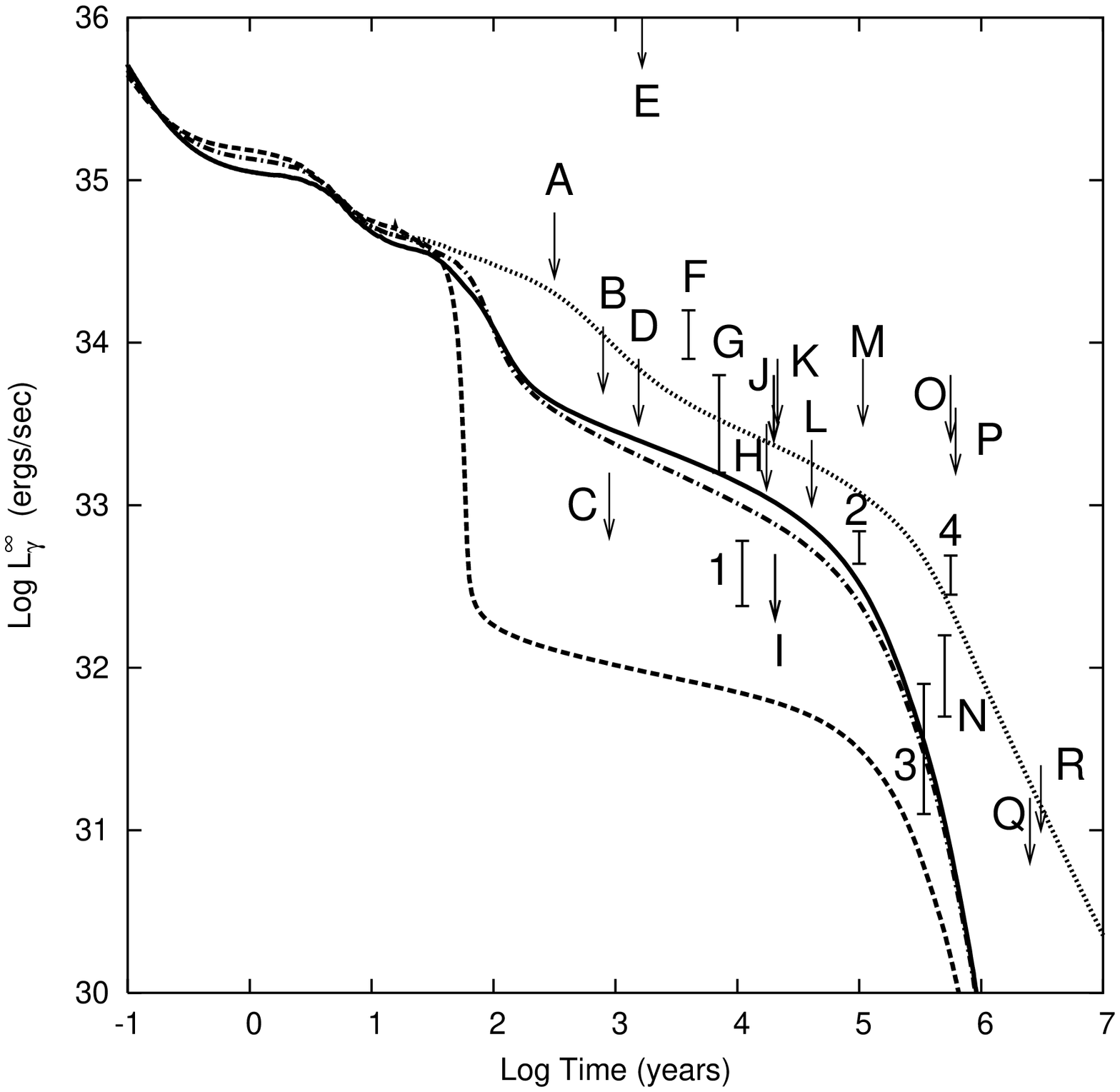}\end{minipage}
\begin{minipage}[t]{0.5\textwidth}
\includegraphics[height=.45\textheight, angle=270]{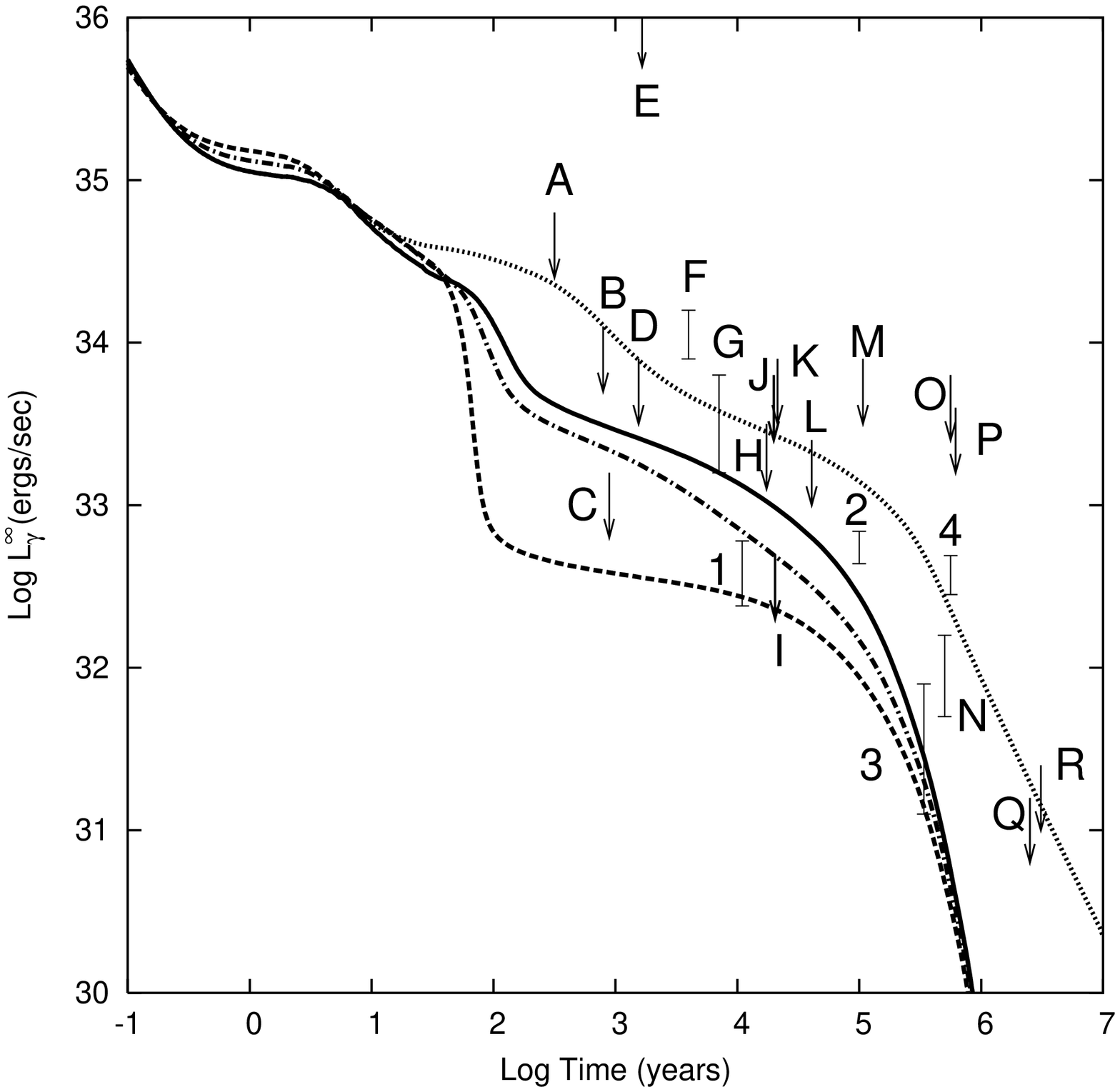}
\end{minipage}

\caption{Thermal Evolution Curves from Tsuruta (2006). In Fig. 1a
(left panel) the dotted and solid curves refer to the standard
cooling of M = 1.4M$_\odot$ neutron stars with and without heating,
respectively, while the dot-dashed and dashed curves are for hyperon
cooling of 1.6 and 1.8M$_\odot$ stars, respectively. In Fig. 1b (the
right panel) the solid, dot-dashed and dashed curves refer to pion
cooling of 1.4, 1.6 and 1.8M$_\odot$ stars, respectively. In the
same figure the dotted curve refers to thermal evolution of a
1.4M$_\odot$ pion star with heating. The vertical bars refer to
temperature detection data with error bars, while the downward
arrows refer to the upper limits. The more accurate detection data
are shown with numbers, for (1) the Vela pulsar, (2) PSR 0656+14,
(3) Geminga, and (4) PSR 1055-52. The rest of the data shown are
more rough estimates. Some of more interesting among these are shown
with letters, as (A) Cas A point source, (B) the Crab pulsar, (C)
PSR J0205+6449 in 3C58, (F) RX J0822-4300, (G) 1E1207.4-5209, (I)
PSR 1046-58, (N) RX J1856-3754, and (R) PSR 1929+10. (\copyright By
permission of the author.)}
\label{CoolingCurves}%
\end{figure}

In a series of papers, Tsuruta et al. has made clear their
conclusions (e.g. Tsuruta et al. 2002; Tsuruta 2004; Tsuruta 2006).
Three points can be summarized,
\begin{enumerate}
    \item Soft EOS, such as BPS should be excluded from neutron star mass
    measurements.\\
    \item Nucleon and kaon DUrca process should be excluded
    especially for the Vela data.\\
    \item Pion cooling is consistent with both observation and
    theory.
\end{enumerate}
A graphical summary is given Figure \ref{CoolingCurves}.

Lattimer \& Prakash (2006) have also made a survey of the neutron
star mass-radius relation. Their result is for Tsuruta's. The
surviving EOSs all supports large masses. Recently, $\ddot O$zel
(2006) has made a compound analysis of the neutron star EXO
0748-676. From its stringent mass radius relation, only the stiffest
EOSs are consistent with the measurement. Conservatively, we
consider both the medium and stiff EOSs from now on. Using point 2
of Tsuruta's conclusion, neutron star cooling could provide us an
independent way of testing the EOS.

Following the above discussion of DUrca process, we can calculate
the critical mass for a specific EOS, above which DUrca process will
turn on in the interior of neutron stars. If the critical mass is
smaller than $1.4\, M_{\odot}$, then the EOS may probably be
excluded. Table \ref{cmass} gives the critical neutron star mass,
calculated for different EOSs.

\begin{table}[t]
  \centering
  \caption{DUrca Critical Mass for Different EOS.}\label{cmass}
  \begin{tabular}{llllll}
    \hline
    q & 0.7 & 0.8 & 0.9 & 1.0 & 1.1 \\ \hline
    FPS & 1.18 & 0.84 & 0.64 & 0.50 & 0.45 \\
    SLy & 1.50 & 1.09 & 0.84 & 0.65 & 0.58 \\
    RMF210 & 1.27 & 1.16 & 1.07 & 0.98 & 0.93 \\
    RMF240 & 1.40 & 1.26 & 1.16 & 1.05 & 0.99 \\
    RMF300 & 1.55 & 1.40 & 1.28 & 1.16 & 1.09 \\
    TNI2u & 1.01 & 0.76 & 0.61 & 0.50 & 0.46 \\
    TNI6u & 1.24 & 0.94 & 0.76 & 0.62 & 0.56 \\
    TNI3u & 1.40 & 1.07 & 0.86 & 0.69 & 0.62 \\
    \hline
  \end{tabular}
\end{table}

\begin{figure}[t]
\centering
\scalebox{0.5}{\includegraphics{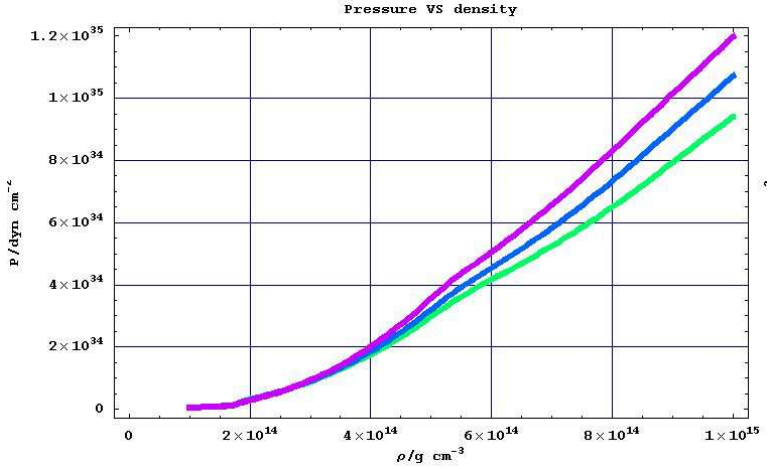}}%

\caption{EOS for RMF210, RMF240, RMF300, from Bottom to Top
Respectively. The behavior of TNI2u, TNI6u, TNI3u is similar.}
\label{RMF}
\end{figure}

The critical mass is given in units of $1\,M_{\odot}$.  FPS
(Friedman Pandharipande Skyrme) and SLy (Skyrme Lyon effective
interaction) are EOSs taken from Haensel \& Potekhin (2004). RMF
means relativistic mean field theory. The quantities 210, 240, 300
are the compression modulus. They are taken from Ma (2002),
Glendenning (1997). The EOSs TNI2u, TNI6u, TNI3u (Three Nucleon
Interaction, u stands for universal inclusion) are taken from
Takatsuka et al.(2006), which Tsuruta's group have used to get their
conclusions. For a comparison of the EOSs, see Figure \ref{RMF}.

When then critical mass is smaller than $1.4\,M_{\odot}$, a normal
neutron star will cool via the rapid DUrca process. Of course, it is
in contradiction to Tsuruta's conclusions. So, soft EOS, such as
FPS, RMF210, TNI2u, may be excluded. Even medium EOS TNI6u is in
danger.

On the other hand, when we choose several stiff EOS, such SLy,
RMF300, these EOS meet the improbability of DUrca process. This time
there is a tendency for smaller q values (Li, Chen 2005). A smaller
q, 0.7, 0.8, etc, satisfies better the astronomical requirement.
This may be tested by further nuclear symmetry energy measurement.
But it may take many years.

Excluding the soft EOSs is consistent with Tsuruta (2006), and
Lattimer \& Prakash (2006). Our result can also be compared with
$\ddot O$zel's (2006). In the case os neutron star cooling, not only
the stiffness of the EOS determines, but also the composition. The
key point is, we present an independent way of testing the EOS, from
the improbability of DUrca process. Following this treatment, we can
separate EOSs more likely from those less likely. Before we come to
the end, there are sereval points to note.
\section{Discussions}
There are four points to note about.
\begin{enumerate}
    \item The presence of muon. When we incorporate muons into our
    consideration, we get a larger minimal proton concentration, but
    smaller critical density (Lattimer et al.
    1991). So the exclusion of soft EOSs is strengthened.

    \item About hyperons. Hyperon DUrca process only add to the more
    effective nucleon DUrca process(Prakash et al. 1992).
    Whether it exists in neutron star interior is still an open question
    (Takatsuka et al. 2006).

    \item The consistent problem. The symmetry energy here is an
    extrapolation of subnuclear density experiment. When we choose
    a fixed form of symmetry energy, we also fix the EOS to some
    degree. Maybe combined with compression modulus data, one can
    make a seemingly more consistent calculation. But it will not
    bother us that, this is an independent way of testing the EOS.

    \item
    The presence of quark matter.
    In this case, things will be more complicated (see general discussion by Pan, Zheng 2007).
    First, the threshold of quark DUrca process is not a simple ingredient fraction.
    Quark-quark interaction must be taken into account (Pethick 1992).
    Second, the cooling scenario in the presence of quark matter have some considerable difference
    with that of a neutron star (Page 2006).
    Moreover, deconfinement heating must be included (Yuan, Zhang 1999; Kang, Zheng 2007).
    Since Tsuruta's exclusion of DUrca process
    is done in the frame of necleon processes, plus pion and kaon condensation.
    If we want to extrapolate our conclusions here, e.g. to that of hybrid stars,
    it will be a systematic project.
    This may be the scope of further studies.

\end{enumerate}

Excluding soft EOSs is a general tendency of neutron star
researches. The preference of smaller q values needs further
studies. As Lattimer said sixteen years ago, \emph{The continuing
attempts to observe thermal radiation from neutron stars will have
important implications for these properties of nuclear matter}. That
is what we try to do here.

\begin{acknowledgements}

The author A is very grateful to Dr. Yu Yun-Wei, for his fruitful
discussions and suggestions. We would like also to thank Dr. Bai
Hua, and Prof. Dai Zi-Gao, for their thoughtful comments.

This research is supported by Chinese National Science Foundation
No.10573011, No.10273006, and the Doctoral Program Foundation of
State Education Commission of China.
\end{acknowledgements}


%
\label{lastpage}
%
\end{document}